\documentclass[twocolumn]{aastex631}
\usepackage{multirow}
\usepackage{natbib}

\def\h{\hskip 0.0 mm}

\def\apjl{{ApJL}}
\def\apjs{{ApJS}}
\def\asec{$^{\prime\prime}$}

\def\e#1{$\times$10$^{#1}$}

\def\hb{H$\beta$}

\def\lambe{$\lambda_{\rm Edd}$}
\def\lax{{$\mathrel{\hbox{\rlap{\hbox{\lower4pt\hbox{$\sim$}}}\hbox{$<$}}}$}}
\def\gax{{$\mathrel{\hbox{\rlap{\hbox{\lower4pt\hbox{$\sim$}}}\hbox{$>$}}}$}}
\def\simlt{\lower.5ex\hbox{$\; \buildrel < \over \sim \;$}}
\def\simgt{\lower.5ex\hbox{$\; \buildrel > \over \sim \;$}}
\def\lum{erg s$^{-1}$}
\def\mbh{{$M_{\rm BH}$}}
\def\micron{{$\mu$m}}

\def\cm2{cm$^{-2}$}

\def\oiii{[\ion{O}{3}]}

\def\lbol{$L_{{\rm bol}}$}

\def\rfe{$R_{\rm Fe\ II}$}
\def\l5100{$L_{5100}$}

\shorttitle{HDD AGNs}
\shortauthors{Son et al.}

\begin{document}

\title{Deficit of Hot Dust in Low-redshift Active Galactic Nuclei}

\author[0000-0002-5346-0567]{Suyeon Son}
\affiliation{Department of Astronomy and Atmospheric Sciences,
Kyungpook National University, Daegu 702-701, Korea}

\author[0000-0002-3560-0781]{Minjin Kim}

\affiliation{Department of Astronomy and Atmospheric Sciences, 
Kyungpook National University, Daegu 702-701, Korea}

\author[0000-0001-6947-5846]{Luis C. Ho}

\affiliation{Kavli Institute for Astronomy and Astrophysics, Peking 
University, Beijing 100871, China}

\affiliation{Department of Astronomy, School of Physics, Peking University, Beijing 100871, China}

\correspondingauthor{Minjin Kim}
\email{mkim.astro@gmail.com}

\begin{abstract}
We assemble a broad-band spectral energy distribution (SED) ranging from optical to mid-infrared of nearby active galactic nuclei at $z < 0.4$. SED fitting analysis is performed using semi-empirical templates derived from Palomar-Green quasars to classify the sample into normal, warm-dust-deficient (WDD), and hot-dust-deficient (HDD) AGNs. Kolmogorov–Smirnov tests reveal that HDD AGNs exhibit, on average higher AGN luminosity than normal and WDD AGNs. HDD fraction, on the other hand, is only weakly correlated with black hole mass and inversely correlated with Eddington ratio.
By fixing the other parameters, we conclude that the HDD fraction is primarily connected with the AGN luminosity. It implies that there is a causal connection between the covering factor of the hot dust component and AGN luminosity, possibly due to the sublimation of the innermost dust or the thickening of the intervening gas in the broad-line region. 
Analysis of the outflow properties traced by the wing of \oiii$\lambda5007$ suggests that outflows may be related to the formation and maintenance of the hot dust component. Finally, we demonstrate through comparison with previous studies that the classification of HDD AGNs requires careful subtraction of the host galaxy light. \end{abstract}


\section{Introduction} 

Active galactic nuclei (AGNs) are crucial to understanding the evolution and formation of supermassive black holes (SMBHs), which grow substantially in the AGN phase \citep[e.g.,][]{soltan_1982, merloni_2008}. As the material swirls onto an SMBH, not only an accretion disk but also additional substructures, such as corona and relativistic jet, can emerge in its vicinity, emitting strong radiation over a wide wavelength range from radio to X-rays. According to the orientation-based unification paradigm, AGNs are classified into type 1 and type 2 depending on whether the dusty torus obstructs the line-of-sight to the central part of the AGN (e.g., accretion disk and broad-line region, \citealt{antonucci_1993, urry_1995}). For example, type 2 AGNs are thought to be covered by the obscuring medium along the line-of-sight. It indicates that torus plays such an important role in the AGN unification model, but its origin and detailed structure are still controversial.

\begin{figure*}[t!]
\centering
\includegraphics[width=0.99\textwidth]{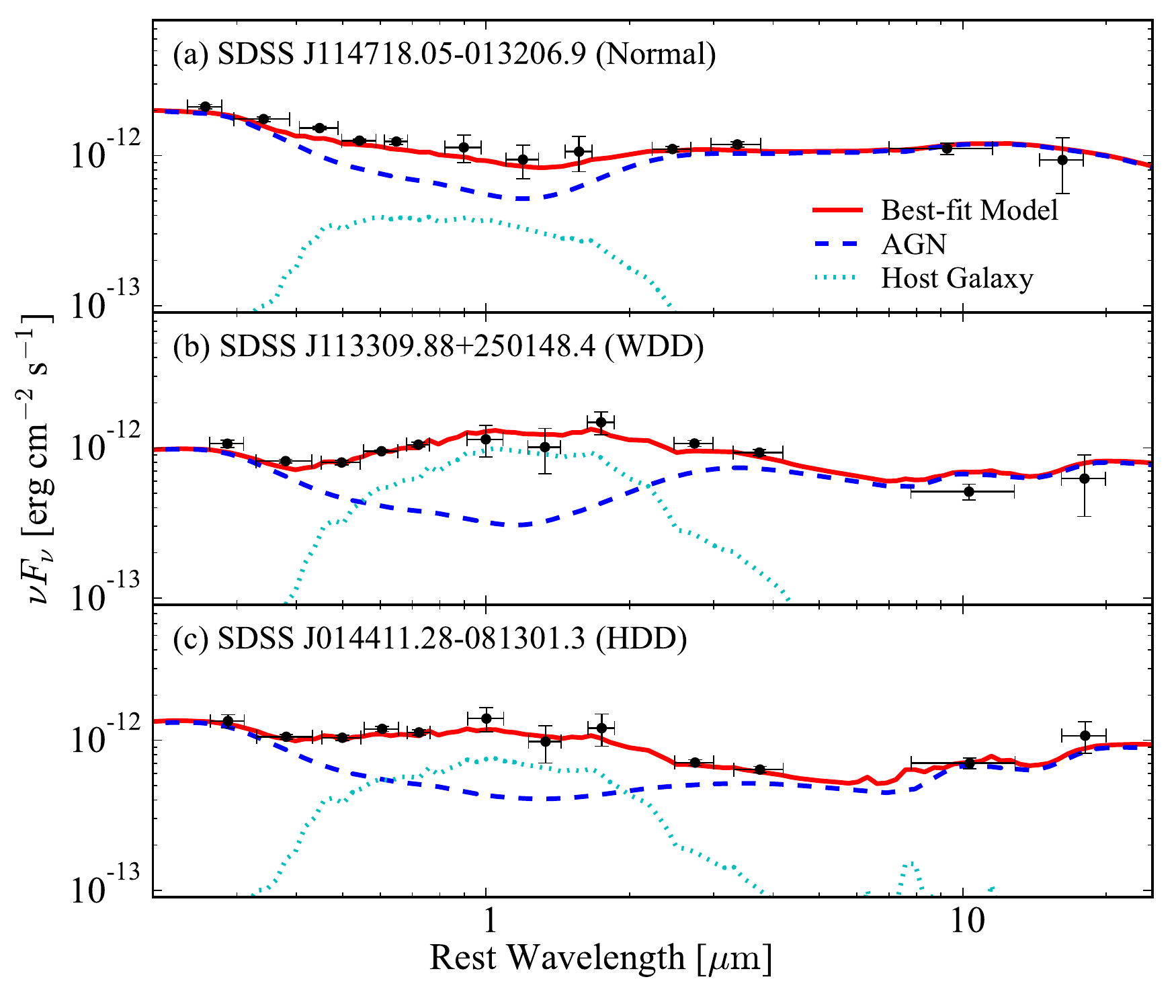}
\caption{
Examples of the SED fitting analysis for normal (a), WDD (b), and HDD (c) AGNs. While black circles represent the observed photometric data, the red solid line denotes the best-fit SED models. Blue dashed and cyan dotted lines denote AGN and host galaxy components for the best-fit model, respectively. 
}
\end{figure*}

As infrared (IR) thermal emission originates from the dust heated by the X-ray/UV light from the accretion disk, the structure of the torus has traditionally been studied with spectral energy distribution (SED) in the IR wavelength,  \citep{rees_1969}. While several studies have suggested that the SED of the torus is well fit with the clumpy clouds rather than the single smooth cloud, an additional hot dust component presumably positioned at the innermost of the torus is still necessary to account for the IR excess \citep[e.g.,][]{nenkova_2002, ogawa_2021}.

Two scenarios have been proposed to explain the physical origin of the torus. (1) The dusty torus is naturally supplied from the host galaxy \citep[e.g.,][]{krolik_1988}. (2) The torus is composed of the dusty material condensed from the AGN outflow \citep[e.g.,][]{elvis_2002}. The two scenarios predict the location and evolution of hot dust in slightly different ways. For example, from the former scenario, hot dust is expected to be located at the sublimation radius because the innermost radius is set where the dust inflow driven by the cloud-cloud collision is sublimated due to the radiation from the accretion disk. Conversely, the latter scenario predicts the hot dust to be positioned outside the sublimation radius because outflowing clouds must reach $\sim1000$ times the initial radius of outflowing clouds for dust condensation (see Section 3 of \citealt{elvis_2002}). Therefore, observational studies to investigate the physical properties of hot dust components will allow us to uncover the origin of the torus, which can be accomplished by reverberation mapping (RM; \citealt{clavel_1989, minezaki_2004, suganuma_2006, koshida_2014, minezaki_2019}) and spatially resolved near-infrared (NIR) interferometric observation \citep{swain_2003, gravity_2020a, gravity_2020b}.

\begin{deluxetable*}{lccccc}
\tablecolumns{6}
\tablenum{1}
\tablewidth{0pc}
\tablecaption{Sample Properties \label{tab:table1}}
\tablehead{
\colhead{\h} &
\colhead{\h Normal} &
\colhead{\h WDD} &
\colhead{\h HDD} &
\colhead{\h Normal+WDD} &
\colhead{\h WDD+HDD} \\
\colhead{\h (1)} &
\colhead{\h (2)} &
\colhead{\h (3)} &
\colhead{\h (4)} &
\colhead{\h (5)} &
\colhead{\h (6)}
}
\startdata
$N$ &\h 550 &\h  991 &\h  547 &\h 1541 &\h 1538 \\
$z$ &\h$0.27\pm0.05$ &\h $0.27\pm0.06$ &\h $0.29\pm0.06$ &\h $0.27\pm0.06$ &\h $0.28\pm0.06$ \\
$\log L_{5100}$&\h$44.16\pm 0.21$ &\h $44.25\pm 0.19$ &\h $44.28\pm 0.21$ &\h $44.22\pm 0.20$ &\h $44.26\pm 0.20$ \\
$\log M_{\rm BH}$ &\h$7.98\pm0.33$ &\h $8.21\pm0.33$ &\h $8.25\pm0.31$ &\h $8.14\pm0.34$ &\h $8.22\pm0.32$ \\
$\log \lambda_{\rm Edd}$ &\h$-0.95\pm 0.33$ &\h $-1.08\pm 0.36$ &\h $-1.10\pm 0.34$ &\h $-1.04\pm 0.35$ &\h $-1.09\pm 0.35$ \\
$\log L_{3.6\mu {\rm m}}$&\h$44.65\pm 0.16$ &\h $44.73\pm 0.16$ &\h $44.53\pm 0.17$ &\h $44.70\pm 0.16$ &\h $44.66\pm 0.17$ \\
$\log L_{12\mu {\rm m}}$&\h$44.40\pm 0.15$ &\h $44.35\pm 0.17$ &\h $44.20\pm 0.18$ &\h $44.37\pm 0.17$ &\h $44.30\pm 0.18$ \\
R$_{\rm Fe\ II}$  &\h$0.63\pm0.32$ &\h $0.51\pm0.28$ &\h $0.35\pm0.26$ &\h $0.55\pm0.29$ &\h $0.45\pm0.28$ \\
\enddata
\tablecomments{
Median values and median absolute deviations for each subgroup.
Col. (1): Properties of the subsamples.
$N:$ the number of subsample; $z:$ redshift;  
$\log L_{\rm 5100}:$ logarithmic 5100 \AA\ monochromatic luminosity in the unit of erg s$^{-1}$;
$\log (M_{\rm BH}):$ logarithmic black hole mass in the unit of $M_{\odot}$;
$\log \lambda_{\rm Edd}:$ Eddington ratio with bolometric luminosity inferred from $L_{\rm 5100}$.;
$\log L_{3.6\mu {\rm m}}:$ Infrared luminosity at 3.6 \micron\ in the unit of erg s$^{-1}$;
$\log L_{12\mu {\rm m}}:$ Infrared luminosity at 12 \micron\ in the unit of erg s$^{-1}$;
$R_{\rm Fe\ II}:$ EW ratio of Fe II to broad \hb. 
Col. (2): Normal AGNs. 
Col. (3): WDD AGNs.
Col. (4): HDD AGNs.
Col. (5): Normal and WDD AGNs.
Col. (6): WDD and HDD AGNs.
}
\end{deluxetable*}

Based on the assumption that the NIR continuum mainly originates from the innermost torus, RM provides a unique opportunity to directly measure the radius of the hot dust component, which is hardly resolved even with high spatial images from {\it Hubble Space Telescope} \citep{antonucci_1985}. Although the inner radius of the torus measured by RM is smaller than that observed by NIR interferometry or predicted by dust size-luminosity relation, the hot dust can still be interpreted as being located in the sublimation radius \citep[e.g.,][]{kishimoto_2007, kawaguchi_2010, kawaguchi_2011}. However, NIR RM and interferometric studies produce inconsistent outcomes for the same target. The RM investigation of NGC 4151 finds that the hot dust radius is proportional to the AGN luminosity \citep{koshida_2009}, probably attributed to a change in dust sublimation radius and may be indicative of a host origin for the dusty torus. Conversely, using NIR interferometry data, \citet{pott_2010} demonstrated that the hot dust radius is independent of the AGN luminosity, suggesting an outflow origin is preferred.

Although the physical properties of hot dust have been widely studied, the hot dust was considered a generic feature of AGN until \citet{jiang_2010} reported two hot-dust-deficient (HDD) quasars at $z\sim6$. Several studies have since explored the properties of HDD quasars, but no consensus has yet been reached on the physical origin of the hot dust deficit. For example, the redshift evolution of HDD fraction was discovered in some studies \citep[e.g.,][]{hao_2010, jun_2013}, but not in others \citep[e.g.,][]{hao_2011, mor_2011}. Furthermore, various studies \citep{jiang_2010, mor_2012, jun_2013} argued that the HDD quasars tend to have lower BH mass than normal quasars, whereas other studies \citep[e.g.,][]{hao_2010, mor_2011, lyu_2017a} failed to find such a trend. The trend for the Eddington ratio (\lambe) is more complicated. \citet{jiang_2010} and \citet{jun_2013} argued that \lambe\ of HDD quasars is higher than that of normal quasars, but \citet{lyu_2017a} reported the contrary. Although it is unclear what causes this discrepancy, insufficient sample sizes, varied HDD selection methods, or/and narrow dynamic ranges of physical properties may make direct comparisons between various studies difficult. 

Based on the above motivation, we investigate the physical properties of a large number of nearby quasar samples in a homogeneous manner to unveil the physical origin of the hot dust component. Section 2 describes sample selection and SED construction using observed data covering the optical to mid-infrared (MIR) wavelength. SED decomposition utilizing semi-empirical AGN templates adopted from \citet{lyu_2017a} and \citet{lyu_2017b, lyu_2018} and its application to SED fit for the AGN classification based on the dust properties are represented in Section 3. Section 4 compares the AGN properties of normal, warm-dust-deficient (WDD), and hot-dust-deficient AGN populations, and presents a discussion of their origins. The summary is presented in Section 5. We adopt the following cosmological parameters: $H_0=100h=67.4$ km ${\rm s}^{-1}$ ${\rm Mpc}^{-1}$, $\Omega_m=0.315$, and $\Omega_\Lambda=0.685$ (\citealt{planck_2020}).

\begin{figure*}[t!]
\centering
\includegraphics[width=0.98\textwidth]{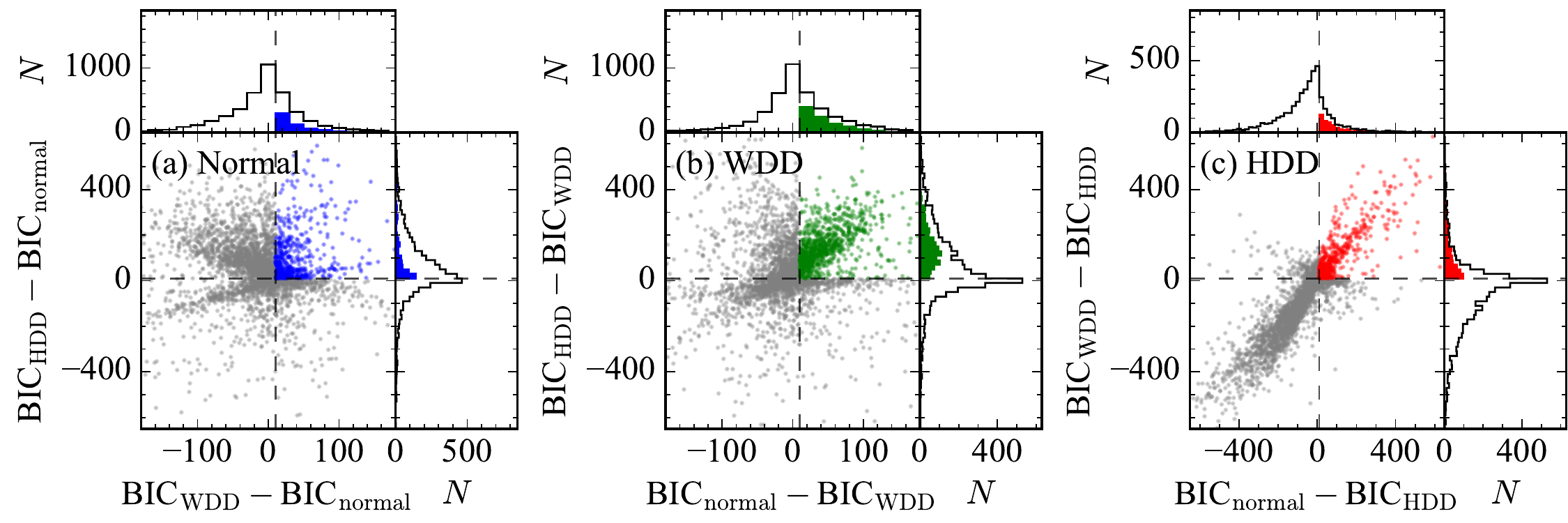}
\caption{
$\Delta$BIC distributions between the subgroups. In each panel, blue, green, and red points represent the sample classified as (a) normal, (b) WDD, and (c) HDD AGNs based on the $\Delta$BIC value ($> 10$) and reduced $\chi^2$ value of the best fit ($\le 10$), respectively. Gray points represent the rest of the sample.
}
\end{figure*}

\section{Sample and Data}

The type 1 AGN sample is initially selected from the 14th data release of the Sloan Digital Sky Survey (SDSS-DR14) quasar catalog \citep{paris_2018}. The redshift limit ($z<0.4$) is set to reduce the effect of cosmic evolution on the torus properties. In addition, this redshift cut is crucial to secure the NIR photometric coverage, which is needed to model the stellar continuum \cite[e.g.,][]{son_2022}. As a result, 7,720 objects are selected. To perform the SED fit, we collect the optical ($ugriz$), NIR ($JHK_s$), and MIR (W1, W2, W3, and W4) photometry from SDSS-DR14, Two Micron All Sky Survey (2MASS) All-Sky Point Source Catalog \citep{2mass_irsa, skrutskie_2006}, and Wide-Field Infrared Survey Explorer (WISE) All-Sky Source Catalog \citep{wright_2010,wise_irsa}, respectively. Therefore, the SED covers the wavelength range of $0.35-22\,\mu{\rm m}$. We employ a radius of 2\asec\ for the cross-match between the catalogs \citep{son_2022}. Finally, 4,355 samples having IR counterparts both in the 2MASS and WISE catalogs are used throughout the paper. The SDSS and 2MASS magnitudes are corrected for the Galactic extinction using a dust reddening map from \citet{schlafly_2011}, based on the reddening law from \citet{fitzpatrick_1999}.

 The spectral measurements from \citet{rakshit_2020} have been adopted in this study to investigate the physical properties of AGNs. BH masses are calculated using a virial method ($M_{\rm BH} \propto v^2r/G$), where the size of the broad-line region ($r$) is inferred from the monochromatic luminosity at 5100 \AA\ (\l5100; e.g., \citealp{bentz_2013}) and the FWHM of \hb\ is used as a surrogate for $v$. In this paper, we employ the BH mass estimator based on the calibration of \citet{ho_2015}. The bolometric luminosity is converted from \l5100\ using a conversion factor of 9.26 \citep{richards_2006}. As a proxy for the accretion rate, we used Eddington ratio ($\lambda_{\rm Edd} \equiv L_{\rm bol}/L_{\rm Edd}$), where $L_{\rm Edd}=1.26 \times 10^{38} \frac{M_{\rm BH}}{M_\odot}$ erg s$^{-1}$.
 Median values with median absolute deviations of various physical properties for each group are shown in Table \ref{tab:table1}.
 
\begin{figure}[t!]
\centering
\includegraphics[width=0.45\textwidth]{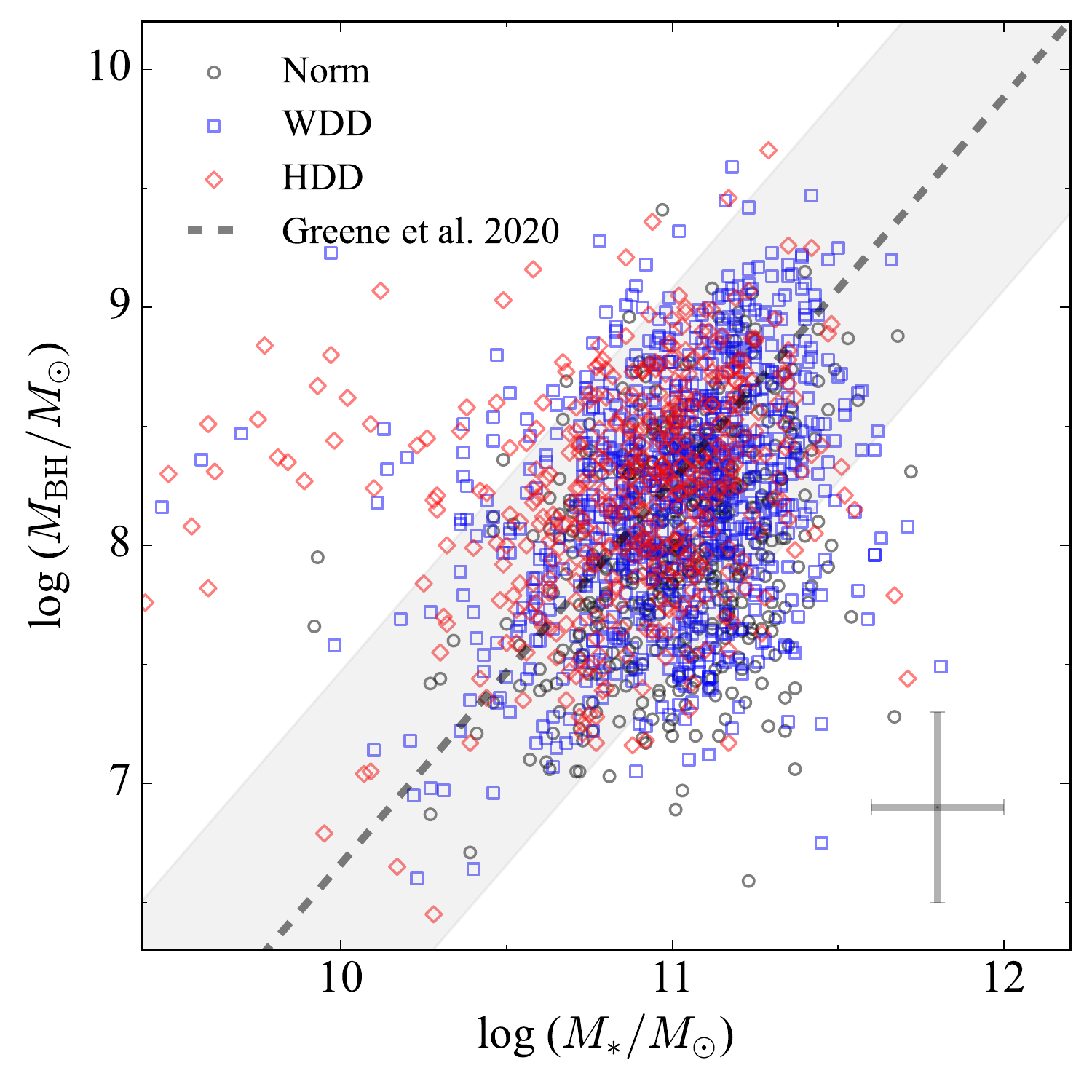}
\caption{
Correlation between BH mass and stellar mass ($M_{*}$) of the host galaxy. The stellar mass is inferred from the best-fit SED model. Black circles, blue squares, and red diamonds represent normal, WDD, and HDD AGNs, respectively. The dashed line denotes the relation of all types of inactive galaxies adopted from \citet{greene_2020}. The shaded area represents the intrinsic scatter ($\sim 0.81$dex) in the $M_{\rm BH}-M_{*}$ relation of normal galaxies. The typical uncertainties of BH mass and stellar mass are shown in the bottom right.}
\end{figure}

\begin{figure}[t!]
\centering
\includegraphics[width=0.45\textwidth]{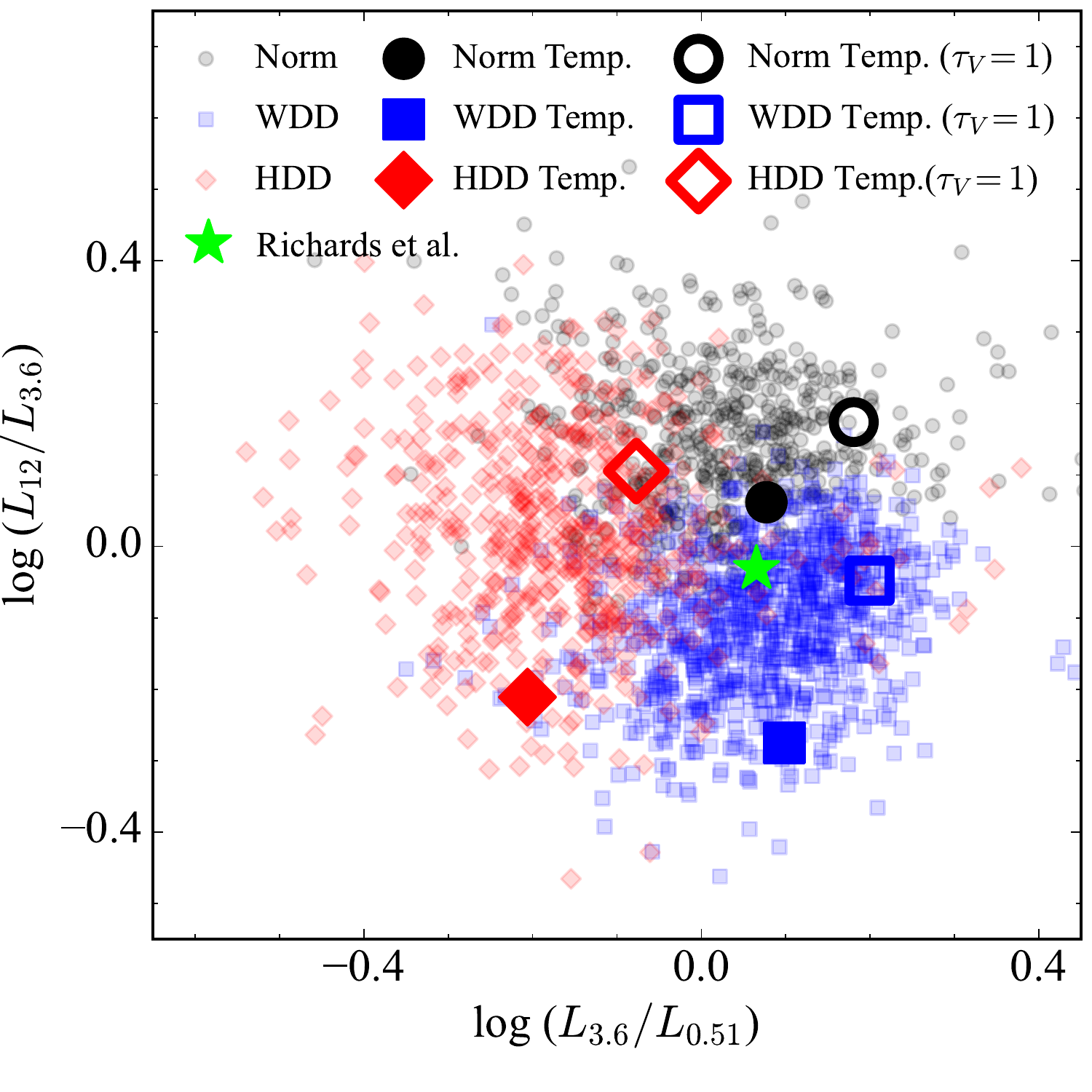}
\caption{
Color-color distribution of our sample AGNs. The horizontal axis is 3.6 $\mu{\rm m}$/0.51 $\mu{\rm m}$ luminosity ratio, and the vertical axis is 12 $\mu{\rm m}$/3.6 $\mu{\rm m}$ luminosity ratio. Symbols are the same as in Figure 3. Large-filled symbols represent the luminosity ratio of template SEDs with no extinction. Large open symbols denote the luminosity ratios of templates SEDs with $\tau_V = 1$ due to the polar dust. The green star represents the luminosity ratios of the QSO template adopted from \citet{richards_2006}.}
\end{figure}

\section{SED Fitting for Torus classification}
\subsection{SED Templates}
While the presence of the hot dust component in the torus can be determined in a variety of ways \cite[e.g.,][]{jiang_2010, hao_2011, mor_2012, jun_2013, lyu_2017a}, SED fitting, which effectively accounts for the contributions from the host galaxy and/or the polar dust component, is used in this study. The semi-empirical templates from \citet{lyu_2017a} are adopted for that purpose, in which the Palomar-Green QSOs \citep{boroson_1992} were classified into three subgroups based on the existence of hot and warm dust components in the NIR/MIR SED: normal, WDD, and HDD AGNs. 
WDD AGNs are defined as those that are deficient in warm dust traced by the thermal emission ranging from $\sim3$ to $\sim20$ $\mu{\rm m}$, while HDD AGNs lack the thermal emission from the additional hot dust ($T\sim1000-2000$ K) around $2-3$ $\mu{\rm m}$ as well as the warm dust emission. More specifically, the dust-deficient AGNs are defined as  those that deviate by at least 0.3 dex from the standard QSO template adopted from \citet{elvis_1994}; see \citealt{lyu_2017a} for more details on the classification. From this classification, \citet{lyu_2017a} generated the representative SED in each subgroup. The stellar continuum from the host galaxy and thermal emission from the polar dust are also included.

For the stellar continuum, the SED template from \citet{lyu_2018} is initially adopted, which is modeled with three components: the 7 Gyr old single stellar population from \citet{bruzual_2003} at $\le 6\,\mu{\rm m}$, MIR composite spectra generated from the Spitzer/IRS spectra of nearby elliptical galaxies without the sign of star formation at $6-20\,\mu{\rm m}$ and a power-law SED ($f_\nu\propto \nu^{1.0}$) above 20 $\mu{\rm m}$. In addition, to account for younger stellar populations commonly discovered in the AGN host galaxies \cite[e.g.,][]{kim_2017, kim_2019, zhao_2021}, we employ the spectral templates from the elliptical galaxy with 2 Gyr old stellar population, S0, Sa, Sb, Sc, and Sd from the SWIRE template library \citep{polletta_2007}. In total, seven templates are used to model the host galaxy. The reddening of the AGN continuum and reprocessed emission due to the polar dust component is taken into account for the SED fit \citep{lyu_2018}. While the polar dust was modeled in detail in \citet{lyu_2018}, it was assumed to follow the power-law density profile ($\rho(r) \propto r^{-0.5}$) and to be composed of large dust grains ($a\ge0.04\ \mu{\rm m}$). The radiative transfer calculation was performed using the code DUSTY (\citealt{ivezic_1997}). Note that the only free parameter is the extinction in $V$ band ($\tau_V$). Finally, the extinction due to the equatorial torus is also considered as some of the sample AGNs exhibit power-law SEDs throughout the optical to MIR region, suggesting non-negligible extinction. To account for this effect, we adopt the extinction law from \citet{fitzpatrick_1999}.

\begin{figure*}[t!]
\centering
\includegraphics[width=0.95\textwidth]{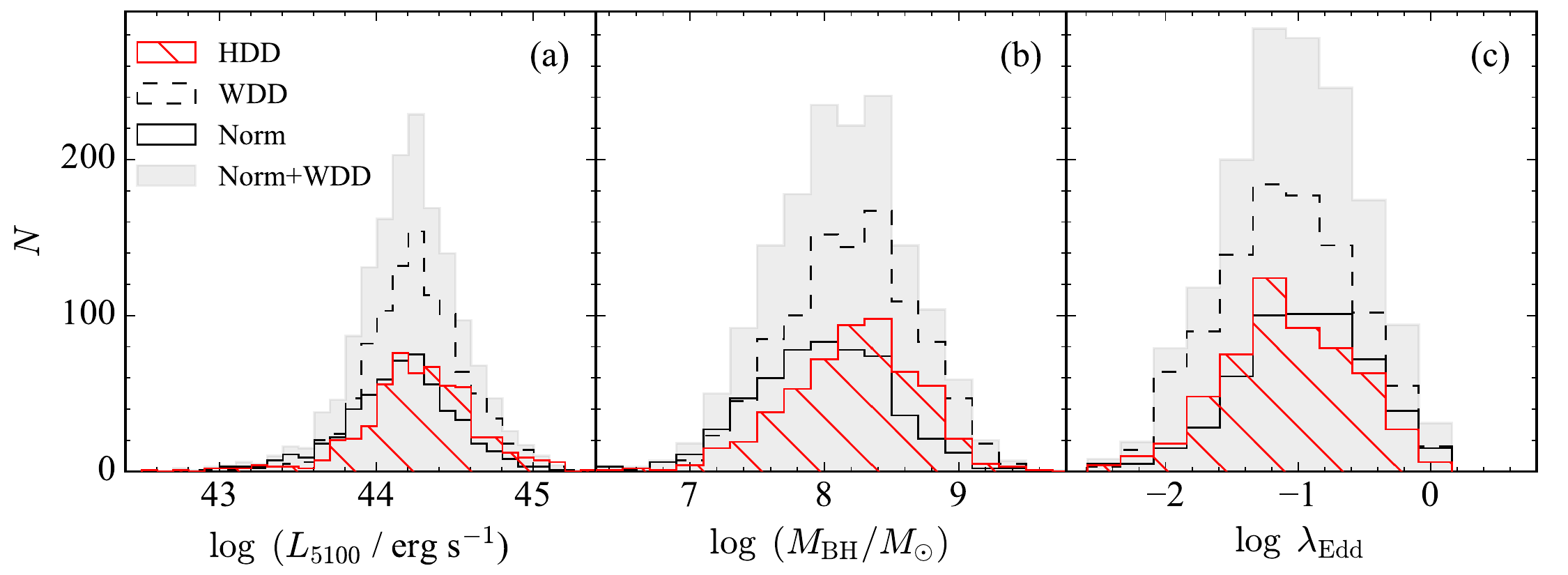}
\caption{
Distributions of AGN properties for each subgroup. (a) Monochromatic luminosity at 5100\AA. (b) BH mass. (c) Eddington ratio.  
}
\end{figure*}

\subsection{SED fitting}
The code LePHARE \citep{arnouts_1999,ilbert_2006} is utilized for the SED fit. The redshift is fixed to the spectroscopic one. During the fit, we utilize all photometric data from $u$-band to W4, where it is available. However, for some objects, a part of NIR/MIR photometry is unavailable in 2MASS or WISE, where only upper limits were given. The relative flux contribution between the AGN SED templates from \citet{lyu_2017a} and the stellar continuum is set at $1.6\,\mu{\rm m}$. We let the stellar contribution at 1.6 $\mu{\rm m}$ range from 1\% to 95\%. We allow the extinction of the stellar continuum to be $A_v=0,\,0.25$, or $0.5$ mag. The extinction of the accretion disk due to the polar dust is forced to be $\tau_v=0,\,0.25,\,0.5,\,0.75$, or $1.0$ to account for the IR contribution from the polar dust. Finally, we let the extinction due to the equatorial torus be $A_V=0.0, 0.3, 0.6, 0.9$, or $1.2$ mag. Examples of the SED fit are displayed in Figure 1.

\subsection{Classification}
As described in \S{3}, we apply the SED fit to each object to find which template out of normal, WDD, and HDD quasars best describe the SED and utilize Bayesian Information Criterion (BIC; \citealt{schwarz_1978}). The BIC is defined as $\mathrm{BIC} \equiv \chi^2 + k \ln N$, where $\chi^2$ is the chi-square of the best-fitting model, $k$ is the number of free parameters, and $N$ is the number of observed data. If the difference in BIC between the two models ($\Delta {\rm BIC}$) is higher than 10, the model with a smaller BIC can be regarded as the best model \citep{liddle_2007}. Finally, we only choose the objects with a reduced $\chi^2 \le 10$ to eliminate the ones with poor SED fits. Figure. 2 displays the distributions of $\Delta {\rm BIC}$ between the subgroups.   

Using the criteria of $\Delta {\rm BIC} > 10$, 550, 991, and 547 objects are classified as normal, WDD, and HDD AGNs, respectively. For the remaining 2267 objects, either we fail to construct a conclusive classification based on the BIC value (i.e., $\Delta {\rm BIC} \le 10$) or their reduced $\chi^2$ is greater than 10 (see Appendix for examples of poor fits). Taken at face value, it is intriguing that the WDD fraction ($\sim47$\%) is significantly higher than that from the previous studies (e.g., $\sim14-17\%$ from PG quasars in \citealt{lyu_2017a}). It is assumed that this is owing to the inclusion of the polar dust in the SED fit, which radiates the emission in the MIR. Note that \citet{lyu_2017a} used the same templates for the classification but did not consider polar dust. As a result, the light contribution from the warm dust can be naturally reduced. To test this hypothesis, we carry out the SED fitting analysis without the polar dust and find that the WDD fraction dramatically decreased to $\sim17$\%, increasing the normal fraction to $\sim65$\%.
It demonstrates that the normal and WDD AGNs can be misclassified from each other mainly due to the degeneracy between the polar dust and warm dust. This trend is also clearly seen in Figure 2, where the $\Delta {\rm BIC}$ distribution between normal and WDD AGNs is centrally concentrated within $\sim10$. Therefore, throughout this study, we do not attempt to distinguish them, instead focusing on the physical difference between two types (i.e., normal and WDD AGNs) and HDD AGNs. On the other hand, the HDD fraction of $\sim$ 26\% in our sample determined from the SED fitting with the polar dust is in good agreement with that of $\sim 15-23$\% from \citet{lyu_2017a}. Note that the HDD fraction based on the SED fit without the polar dust is $\sim17\%$ in our sample. Throughout this study, we employ the classification based on the SED fitting analysis that includes the polar dust.

\section{Discussions}

\subsection{Reliability of the SED Analysis}
To test the reliability of the SED fitting analysis, we estimate the stellar mass of the host galaxies from the fitting results. The majority of SED templates for the host galaxy used in the analysis are empirically determined. Therefore, we estimate the stellar mass based on the mass-to-light ratio ($M/L$) in the $H$ band. As the host spectrum peaks around 1.6 $\mu$m in $F_\nu$, we first compute the $H$-band magnitude from the host SED in the best-fit model. $M/L$ at the $H$-band is inferred using the equation in \citet{bell_2003} associated with $B$-$R$ color, which is determined by its host morphology from the best-fit model \citep{fukugita_1995}. Note that we also compute the stellar mass based on the SED fit without the polar dust and find that the stellar mass of the host galaxies agrees with that from the SED fit with the polar dust within 0.02 dex. The comparison between the BH mass and the total stellar mass of the host galaxy reveals that the relation between two quantities of our sample agrees well with that of normal galaxies \citep{greene_2020}, suggesting that our SED fitting analysis is reliable (Fig. 3). The BH mass estimate based on the virial method is known to have a typical uncertainty of $\sim0.4$ dex (e.g., \citealt{vestergaard_2006}), whereas stellar mass inferred from the optical color can be estimated with an accuracy of $\sim0.2$ dex (e.g., \citealt{bell_2003}).

We additionally utilize the optical and MIR color information to examine whether our classification agrees well with the color-based selection. After some experiments, we find that three subsamples can be most distinctive in the plane of 3.6 $\mu {\rm m}$ to 0.51 $\mu {\rm m}$ flux ratio and 12 $\mu {\rm m}$ to 3.6 $\mu {\rm m}$ flux ratio (Fig. 4). However, there exist substantial overlaps among the subsamples, possibly due to the light contribution from the host galaxy as well as the effect of the polar dust. For example, as illustrated in Figure 4, the extinction with $\tau_V=1$ owing to the polar dust results in dramatic changes in the optical and MIR color. It implies that careful treatment of both components may be crucial for the torus classification.

\begin{figure*}[t!]
\centering
\includegraphics[width=0.95\textwidth]{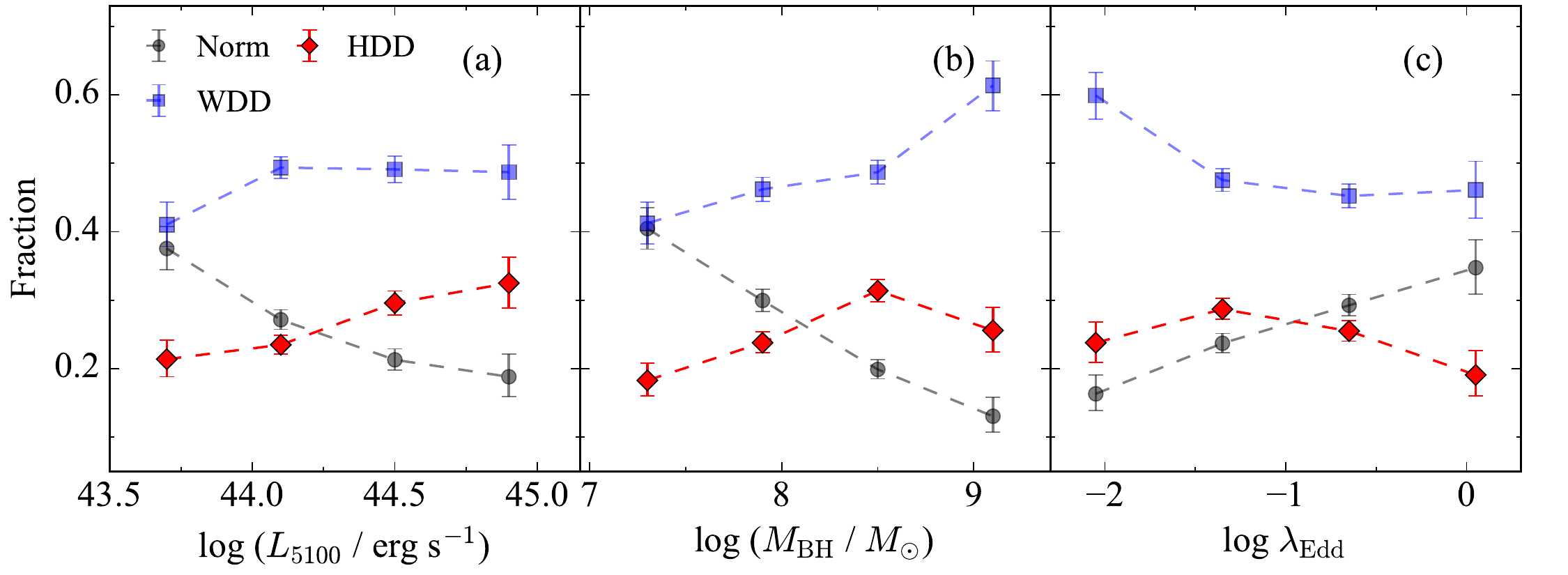}
\caption{
HDD fraction (red diamonds) as a function of $L_{\rm 5100}$ (left), BH mass (middle), and \lambe\ (right). The error bar denoted a 68\% confidence level determined from the Jeffreys confidence interval \citep{brown_2001}. Fractions for normal and WDD AGNs are denoted by black circles and blue squares, respectively.}
\end{figure*}

\subsection{Dependence on AGN Properties}
We analyze the AGN properties in the subgroups to determine the nature of the hot dust deficiency in nearby AGNs. Specifically, we perform Kolmogorov–Smirnov (K-S) test to examine whether HDD AGNs are distinguishable from normal or WDD AGNs. While numerous physical parameters are examined, we emphasize crucial physical parameters potentially responsible for the HDD phenomena. As summarized in Table \ref{tab:table2} and Figure 5, \l5100\ and \mbh\ of HDD AGNs and those of normal and WDD AGNs appear not to be drawn from the same distribution with a significance level of $< 0.1\%$ (i.e., $p-$value $< 0.001$). In particular, HDD AGNs tend to have a larger \l5100\ than normal and WDD AGNs. However, the distributions of the other two parameters (\lambe\ and \mbh) are only distinguishable between normal and HDD AGNs. In contrast, such distinctions between WDD and HDD AGNs are not apparent with $p-$values $> 0.05$, which implies that the null hypothesis that the two distributions are drawn from the same parent population cannot be rejected. Finally, the K-S test reveals that the distribution of \lambe\ for normal AGNs and both WDD and HDD AGNs may not be drawn from the same at a confidence level greater than 99.9\%. 
As described in \SS{3.3}, the distinction between normal and WDD AGNs is somewhat ambiguous due to the degeneracy between MIR emission from warm dust and polar dust, revealing that we caution against overinterpreting physical distinctions between normal and WDD AGNs.

\begin{figure*}[t!]
\centering
\includegraphics[width=0.95\textwidth]{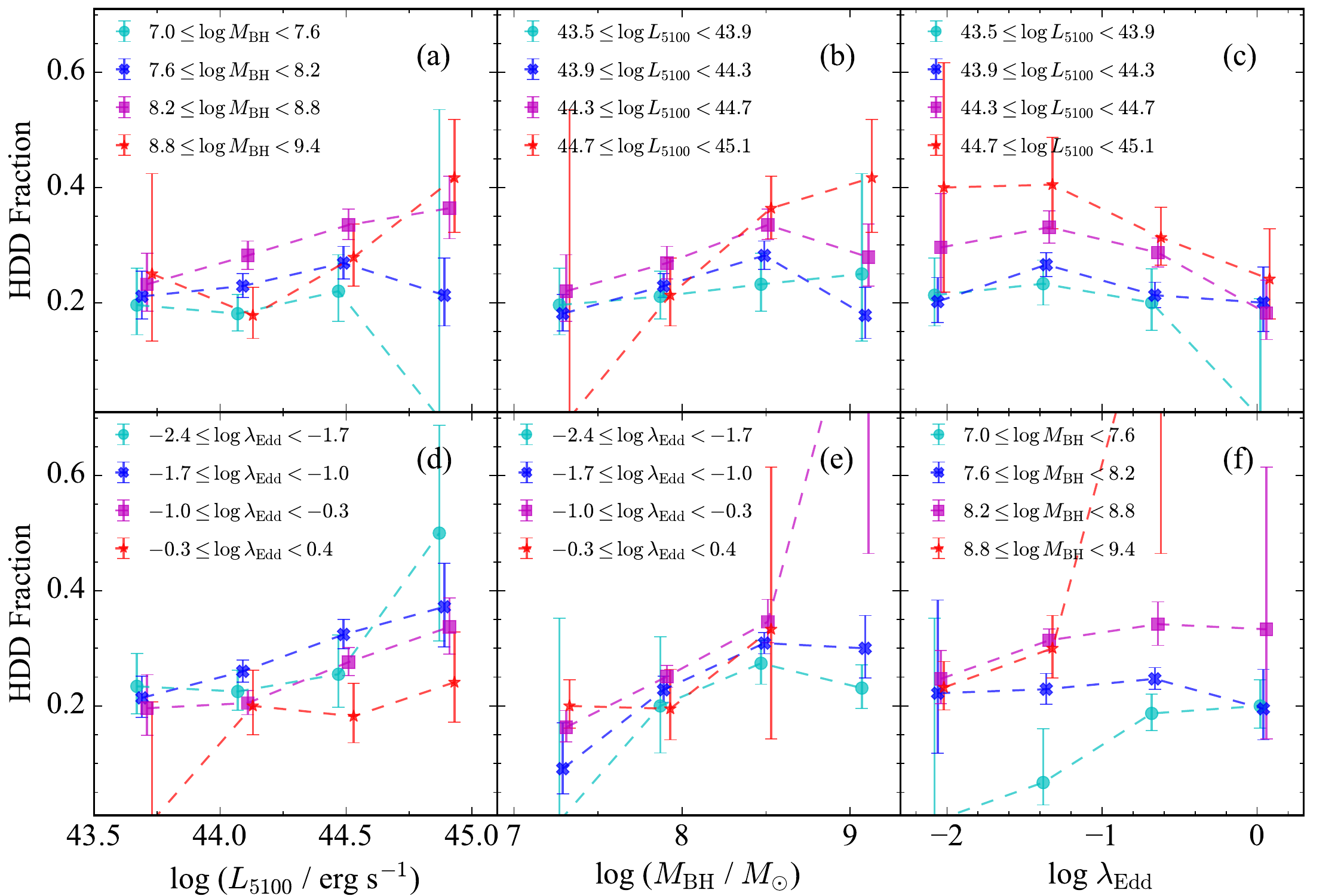}
\caption{
HDD fraction as a function of AGN physical parameters for subgroups. 
We show HDD fraction vs. 5100 \AA\ luminosity, BH mass, and Eddington ratio from left to right. The sample is divided into four subgroups based on the BH mass (a,f),
5100 \AA\ luminosity (b,c) and Eddington ratio (d,e). 
}
\end{figure*}

\begin{deluxetable*}{lccccccccccccccc}
\tablecolumns{15}
\tablenum{2}
\tablewidth{0pc}
\tablecaption{K-S Test Results \label{tab:table2}}
\tablehead{
\colhead{\h} &
\multicolumn{2}{c}{\h Normal vs. WDD} &
\colhead{\h} &
\multicolumn{2}{c}{\h Normal vs. HDD} &
\colhead{\h} &
\multicolumn{2}{c}{\h WDD vs. HDD} &
\colhead{\h} &
\multicolumn{2}{c}{\h Normal vs. WDD+HDD} &
\colhead{\h} &
\multicolumn{2}{c}{\h Normal+WDD vs. HDD} \\
\cline{2-3}
\cline{5-6}
\cline{8-9}
\cline{11-12}
\cline{14-15}
\colhead{\h} &
\colhead{\h $\Delta$} &
\colhead{\h $p$} &
\colhead{\h} &
\colhead{\h $\Delta$} &
\colhead{\h $p$} &
\colhead{\h} &
\colhead{\h $\Delta$} &
\colhead{\h $p$} &
\colhead{\h} &
\colhead{\h $\Delta$} &
\colhead{\h $p$} &
\colhead{\h} &
\colhead{\h $\Delta$} &
\colhead{\h $p$} 
\\
\colhead{\h (1)} &
\colhead{\h (2)} &
\colhead{\h (3)} &
\colhead{\h} &
\colhead{\h (4)} &
\colhead{\h (5)} &
\colhead{\h} &
\colhead{\h (6)} &
\colhead{\h (7)} &
\colhead{\h} &
\colhead{\h (8)} &
\colhead{\h (9)} &
\colhead{\h} &
\colhead{\h (10)} &
\colhead{\h (11)}  
}
\startdata
$z$ &\h$+0.00$ &\h 0.573 &\h  &\h $-0.01$ &\h 0.007 &\h  &\h $-0.02$ &\h 0.012 &\h  &\h $-0.00$ &\h 0.157 &\h  &\h $-0.02$ &\h 0.008 \\
$\log L_{5100}$&\h$-0.09$ &\h $<0.001$ &\h  &\h $-0.12$ &\h $<0.001$ &\h  &\h $-0.03$ &\h 0.098 &\h  &\h $-0.09$ &\h $<0.001$ &\h  &\h $-0.06$ &\h $<0.001$ \\
$\log M_{\rm BH}$ &\h$-0.23$ &\h $<0.001$ &\h  &\h $-0.27$ &\h $<0.001$ &\h  &\h $-0.04$ &\h 0.426 &\h  &\h $-0.24$ &\h $<0.001$ &\h  &\h $-0.11$ &\h $<0.001$ \\
$\log \lambda_{\rm Edd}$  &\h$+0.13$ &\h $<0.001$ &\h  &\h $+0.15$ &\h $<0.001$ &\h  &\h $+0.02$ &\h 0.567 &\h  &\h $+0.14$ &\h $<0.001$ &\h  &\h $+0.06$ &\h 0.189 \\
\enddata
\tablecomments{
Results of two-sample K–S tests. 
Col. (1): Properties of the subsamples. Same as Table 1.
Col. (2): Difference in the median values for normal AGNs and WDD AGNs. $\Delta\equiv M_{\rm norm}-M_{\rm WDD}$, where $M_{\rm norm}$ and $M_{\rm WDD}$ are the median values for normal AGNs and WDD AGNs, respectively.
Col. (3): $p$-value for normal AGNs against WDD AGNs.
Col. (4): Difference in the median values for normal AGNs and HDD AGNs. 
Col. (5): $p$-value for normal AGNs against HDD AGNs.
Col. (6): Difference in the median values for WDD AGNs and HDD AGNs.
Col. (7): $p$-value for WDD AGNs against HDD AGNs.
Col. (8): Difference in the median values for normal AGNs and WDD+HDD AGNs.
Col. (9): $p$-value for normal AGNs against WDD+HDD AGNs.
Col. (10): Difference in the median values for normal+WDD AGNs and HDD AGNs.
Col. (11): $p$-value for normal+WDD AGNs against HDD AGNs.
}
\end{deluxetable*}
\subsection{What Makes Hot Dust Deficiency?}

From the K-S tests, we argue that three physical parameters (\l5100, \mbh, and \lambe) might be responsible for the hot dust deficiency. To quantify the correlations, we estimate the fraction of HDD as a function of those parameters (Fig. 6). The HDD fraction increases considerably as AGN luminosity increases. A Pearson correlation coefficient ($r$) of 0.98 with a $p-$value of $\sim0.018$ between the HDD fraction and AGN luminosity implies that the correlation is significant. The results are not very sensitive to the exact choice of binning. However, its correlation with BH mass and Eddington ratio is more complicated in that HDD fraction peaks around \mbh\ $\sim10^{8.5}\, M_\odot$ and \lambe\ $\sim0.1$ and slightly decreases with increasing \mbh\ and \lambe, respectively. Concerning the dependence on \mbh\ and \lambe, we suspect it mainly comes from the difference between normal and HDD AGNs; there is no significant difference in \mbh\ and \lambe\ distributions between WDD and HDD AGNs (Tables 1 and 2). Nevertheless, it is interesting to note that HDD fraction declines with Eddington ratio at a low Eddington ratio (\lambe\ $\sim 0.01$), while it is only weakly dependent on Eddington ratio at \lambe\ $> 0.05$. This will be further discussed later in this section. As a result, we conclude that \l5100\ could be a primary driver of the HDD phenomenon. To investigate this scenario further, we divide the sample into subgroups based on \mbh\ and \lambe\ (Fig. 7). This experiment reveals that the dependency on \l5100\ still persists even with fixing other parameters.

\l5100\ is known as a good tracer of AGN bolometric luminosity. Therefore, its correlation with HDD fraction reveals that the HDD fraction increases with increasing AGN luminosity. The dependence on the AGN luminosity can be interpreted as the covering factor of the hot dust component decreasing with increasing the AGN luminosity. The anti-correlation between the IR-to-optical flux ratio and the AGN luminosity was previously recognized \citep[e.g.,][]{maiolino_2007, treister_2008}. To explain this trend, the receding torus model has often been invoked, in which the inner radius of the torus grows with AGN luminosity, presumably due to the sublimation \citep[e.g.,][]{lawrence_1991, simpson_2005, honig_2007}. From the MIR variability investigations of nearby AGNs, \citet{son_2022} also argued that the covering factor of the hot dust potentially decreases as the AGN luminosity increases. 

Alternatively, the bowl-shaped torus \citep{gaskell_2007, kawaguchi_2010, kawaguchi_2011, goad_2012}, driven by the anisotropic accretion disk emission, can also be attributed to this dependence. In this model, optically thick gas in the broad-line region (BLR) lies along the bowl-shaped sublimation rim, obstructing the accretion disk emission towards part of the torus close to the equatorial plane. Consequently, most of the hot dust emission comes from the top edge of the torus, and the covering factor of hot dust is dependent on how much BLR gas covers the sublimation rim (i.e., the height ratio of BLR gas to torus). As expected from this model, the hot dust covering factor \citep[$\sim 0.07-0.1$ on average;][]{landt_2011, mor_2011} is known to be substantially smaller than that of the ordinary torus \citep[$\sim0.4$ on average; e.g.,][]{sanders_1989}. Previous studies demonstrated that the height of the torus depends on AGN luminosity as $h_\mathrm{torus} \propto L^{\xi}$, where $\xi$ ranges from 0.23 to 0.4 \citep{cao_2005,simpson_2005,lusso_2013}. On the other hand, gas tends to puff up more easily than dust when the gravitational potential decreases with distance from the BH. For example, \citet{ramolla_2018} have inferred that the height of BLR gas depends on the equatorial radius ($h_\mathrm{BLR} \propto R^{1.5}$) at a given turbulent velocity and viscosity parameter. From the size-luminosity relation of the BLR ($R\propto L^{0.5}$), we can infer that the height of the BLR increases more rapidly with $L$ than that of the torus ($h_\mathrm{BLR} \propto L^{0.75}$ vs. $h_\mathrm{torus} \propto L^{0.23-0.4}$). Therefore, the thickening of the BLR due to the high AGN luminosity can decrease the covering factor of the hot dust component.  

A similar trend was found in \citet{mor_2011}, in which they reported that the covering factor of hot dust is strongly correlated with \lbol\ but not with \mbh\ or \lambe. However, contrary to our result, previous studies about HDD quasars have shown that the HDD fraction is independent of AGN luminosity \citep[e.g.,][]{jiang_2010, hao_2010, hao_2011, mor_2011, lyu_2017a}. 
This discrepancy is partly because some previous studies were conducted with luminous AGNs, which can be inadequate to identify the luminosity dependence of HDD fraction. For example, \citet{jiang_2010, hao_2011, mor_2011} focused on the AGNs with $L_{\rm bol} \gtrsim10^{45.5-46.5}$ \lum, which is $\sim10$ times higher than AGN luminosity of our sample. In addition, studies with AGNs over a wide range of redshift found no evidence of dependence on AGN luminosity, possibly due to the difficulty of disentangling the luminosity dependence of HDD fraction from cosmological effects (e.g., \citealt{hao_2010, hao_2011}). Interestingly, nearby AGNs that cover a wide range of AGN luminosity exhibit an inverse correlation between the covering factor of the hot dust component and AGN luminosity (\citealt{mor_2012}), consistent with our finding.        

\begin{figure}[t!]
\centering
\includegraphics[width=0.45\textwidth]{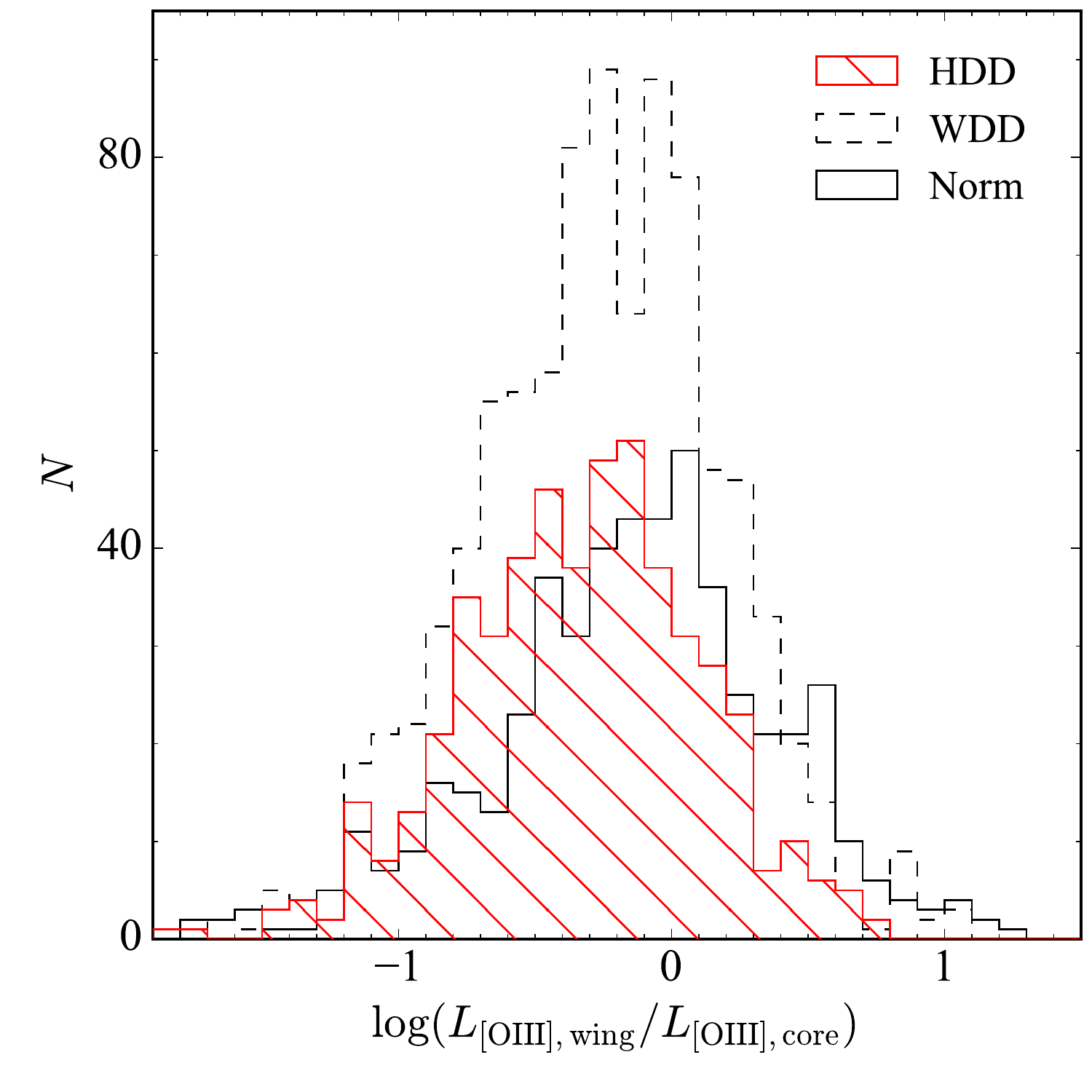}
\caption{
The ratio of \oiii\ luminosity of the wing component to that of the core component for different subsamples. The symbols are the same as in Figure 5.}
\end{figure}

On the contrary, recent studies suggested that the covering factor of the dust can be independent of the AGN luminosity but relatively sensitive to the Eddington ratio \citep[e.g.,][]{stalevski_2016, buchner_2017, ezhikode_2017, ricci_2017c}. For example, \citet{lyu_2017a} argued that only the \lambe\ distribution was distinguished between HDD and normal AGN while the distributions of \lbol\ and \mbh\ for HDD and normal AGNs may be drawn from the same parent population. The dependence on the Eddington ratio can be explained by the relative strength of radiation pressure compared to the gravitational potential of SMBH \citep[e.g.,][]{fabian_2009, wada_2015}. According to \citet{ricci_2017c}, the obscuring material in the circumnuclear region around SMBHs can be easily blown away owing to the radiation pressure. Interestingly, the radiation pressure effectively clears the obscuring material only at \lambe\ $\gtrsim 0.03$, and the hot dust in the vicinity of the accretion disk with a smaller \lambe\ than 0.03 may persist due to the low radiation pressure \citep[e.g.,][]{fabian_2009, ricci_2017c, venanzi_2020}. This prediction is in good agreement with our finding that HDD fraction declines at a low Eddington ratio (\lambe\ $\sim 0.01$) and is weakly correlated with the Eddington ratio $\sim 0.01-0.04$. On the other hand, we find the gradual decline of HDD fraction at a high Eddington ratio (\lambe\ $\sim 0.04-1$), coincident with a critical Eddington ratio (\lambe\ $\sim 0.3$) at which the standard thin disk \citep{shakura_1973} is known to turn into the slim disk \citep{abramowicz_1988}. This decline can be explained by reduced radiation pressure on the torus attributed to anisotropic radiation from the geometrically thick inner structure of the slim disk, thereby increasing the covering factor of the dusty torus. Similar to our result, \citet{zhuang_2018} reported that the torus half-opening angle first decreases and then increases with increasing \lambe\ and argued that it could be caused by the geometrical change of the accretion disk with Eddington ratio.

To examine whether the outflow is responsible for the hot-dust deficiency, we independently investigate the presence and strength of the outflow using the \oiii$\lambda5007$ emission line. The \oiii\ line is commonly modeled with two components \citep[e.g.,][]{heckman_1981, nelson_1996, greene_2005a}. In general, a core component of relatively narrow line width is governed by the gravitational potential of the host galaxy, while non-gravitational motions, such as those driven by an outflow \citep[e.g.,][]{crenshaw_2010, mullaney_2013}), produce broader wings. From the spectral measurements of \citep{rakshit_2020}, we find that fractions of objects with an \oiii\ wing are $94.4\pm1.0\%$, $92.9\pm1.0\%$, and $92.4\pm1.2\%$ for normal, WDD, and HDD AGNs, respectively, tentatively suggesting that outflows may be marginally more common in normal AGNs. To estimate the relative strength of the outflow, we compute the luminosity ratio of the wing relative to the core ($L_{\rm [O\ III], wing} / L_{\rm [O\ III], core}$). The medians and their absolute deviations of $\log (L_{\rm [O III], wing} / L_{\rm [O III], core})$ are $-0.15\pm0.33$, $-0.23\pm0.31$, and $-0.31\pm0.29$ for normal, WDD, and HDD AGNs, respectively. In addition, the K-S test between normal+WDD AGNs and HDD AGNs yields a $p-$value $< 0.001$, which again suggests that the outflow in normal and WDD AGNs can be stronger than that in HDD AGNs (Figure 8). These findings appear to disagree with the notion that the outflow suppresses the hot dust component. Instead, the outflow may enhance the formation of hot and warm dust. This trend is consistent with the ``radiation-driven fountain model'', where the dusty torus is formed and maintained by the outflow from the accretion disk \citep[e.g.,][]{elitzur_2006, wada_2015}.

The ratio (\rfe) of equivalent width (EW) of Fe II measured within 4434--4684 \AA\ to that of broad \hb\ correlates with eigenvector 1, which may be determined primarily by the Eddington ratio \citep[e.g.,][]{boroson_2002, shen_2014}. We find that \rfe\ tends to be marginally larger for normal AGNs than HDD AGNs (Tab. \ref{tab:table1}). Therefore, its correlation with HDD fraction indirectly reveals that the HDD fraction increases with decreasing the Eddington ratio, consistent with the finding in this study. Interestingly, a similar trend was found in other studies: \citet{shen_2014} reported that $r-$W1 color, which is an approximate indicator of the relative strength of the hot dust emission (Fig. 4), is correlated with \rfe.   

Finally, similar to the trend for \lambe, we also find that the HDD fraction increases with \mbh\ at \mbh\ $\lesssim 10^{8.5}\,M_\odot$ and decreases with \mbh\ at \mbh\ $\gtrsim 10^{8.5}\,M_\odot$, although the decrease in the high-mass end is statistically meaningful only at the $2\sigma$ confidence. As shown in Figure 7, even with fixing the other parameters, this trend still remains the same, contradicting previous studies which claimed HDD AGNs tend to have a lower \mbh\ \citep[e.g.,][]{jiang_2010, mor_2012, jun_2013}. This discrepancy may come from the fact that previous studies focused on the luminous and therefore massive AGNs than our study. The physical origin of the declines in the HDD fraction at low and high BH masses is unclear and may be addressed in future studies with the enlarged sample.

\begin{figure}[t!]
\centering
\includegraphics[width=0.45\textwidth]{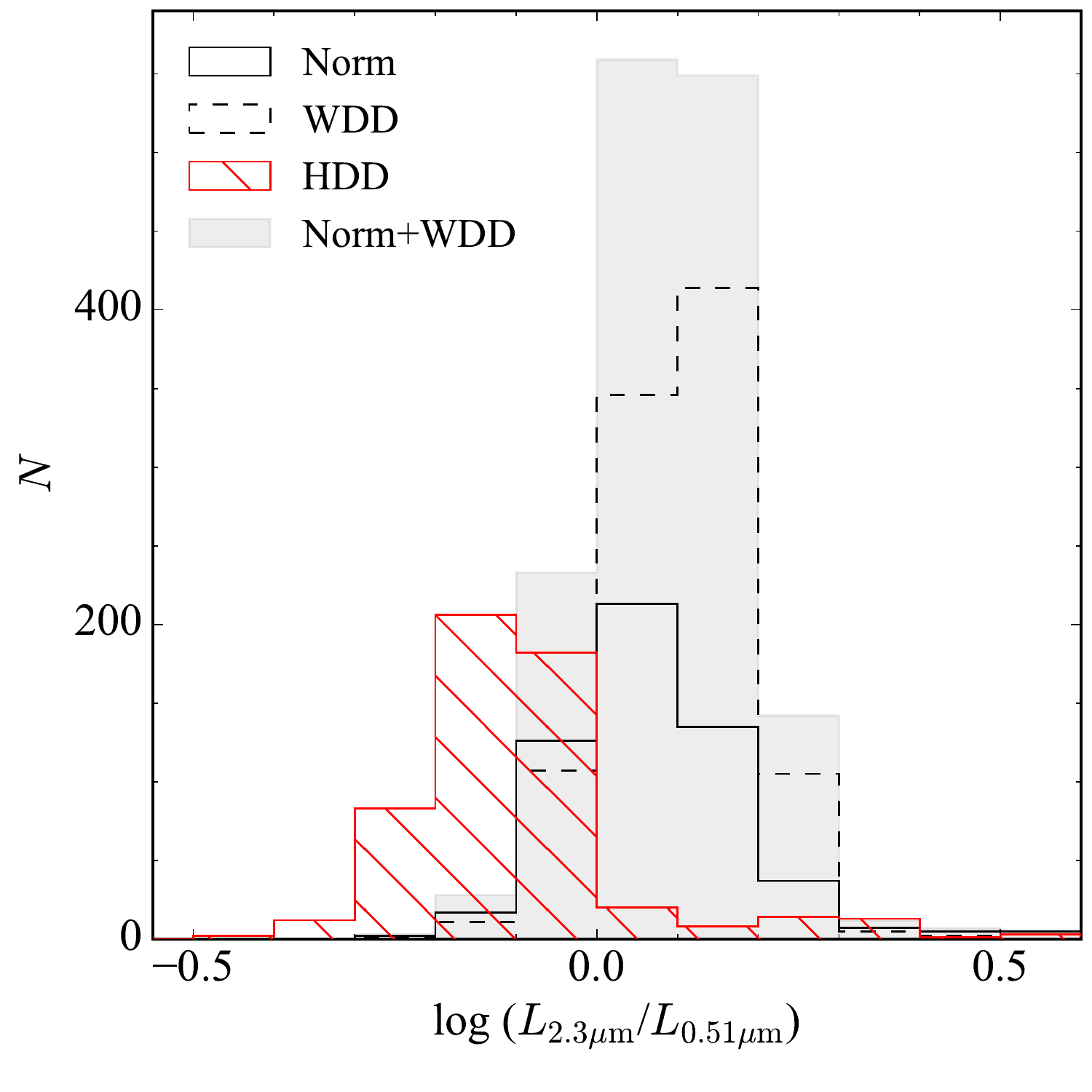}
\caption{
Distributions of the flux ratio of 2.3 \micron\ to 0.51 \micron. The symbols are the same as in Figure 5. 
}
\end{figure}

\subsection{Comparison of Selection Method with Previous Studies}
The dust-deficient AGNs were identified in various ways, which may induce bias in the sample selection between studies. Therefore, it is worthwhile to compare our selection method with previous studies. One of the most common approaches is to use the optical-IR color \citep[e.g.,][]{hao_2010, jiang_2010, jun_2013}. For example, \citet{jun_2013} employed 2.3 \micron\ to 0.51 \micron\ flux ratio to identify the HDD QSOs, whereas 3.4 \micron\ flux density was instead used to trace the covering factor of the hot dust in \citet{jiang_2010}. We adopt the selection method from \citet{jun_2013} for the comparison with our study. 

The luminosities at 0.51 and 2.3 \micron\ ($L_{2.3}$ and $L_{0.51}$) are calculated in the rest-frame, using the linear interpolation of the observed broadband SED. Note that no assumption about the best-fit model is made in this calculation. As expected, we find that HDD AGNs had reduced $L_{2.3}/L_{0.51}$ ratios (Fig. 9). However, its distribution partially overlaps with the other two populations (normal and WDD AGNs). Furthermore, $\log (L_{2.3}/L_{0.51})$ of HDD AGNs is significantly greater than the criterion ($-0.5$) for selecting the HDD in \citet{jun_2013}. It should be noted that \citet{jun_2013} dealt with very luminous AGNs ignoring the flux contribution from the host galaxy. In contrast, the sample in our study is substantially fainter than that in \citet{jun_2013}, indicating that our sample may suffer from the host contribution. Therefore, careful treatment of the host light is required for HDD selection, and the simplified color-cut selection may not be reliable, especially for low-luminosity AGNs.

\section{Summary}
To understand the physical origin of the hot dust deficiency in AGNs, we utilize the broadband spectral energy distribution of low-$z$ AGNs at $z<0.4$ selected from the SDSS quasar catalog. We adopt the SED templates generated from Palomar-Green quasars to classify the sample based on the hot dust in the SED. We perform the SED fitting analysis with careful treatments on the stellar continuum from the host galaxy and extinction and emission due to the polar dust. From these experiments, we find the following results:

\begin{itemize}
\item 26\% of the sample is unlikely to exhibit hot and warm dust components (i.e., hot-dust-deficient AGNs), while 47\% of the sample is classified as warm-dust-deficient (WDD) AGNs.

\item Hot-dust-deficient (HDD) AGNs tend to have larger \l5100\ than normal and WDD AGNs.

\item HDD fraction first increases in a low Eddington ratio and then decreases in a high Eddington ratio with Eddington ratio. A similar up-and-down trend with BH mass is also detected. However, as this trend is statistically significant only at the level of 2$\sigma$, it needs to be further tested with a larger sample. 

\item Dependency on \l5100\ is consistent with the receding torus model, in which the covering factor of the hot dust is anti-correlated with AGN luminosity, possibly due to the sublimation of the innermost torus. In addition, this trend can also be explained by enlarging the height of the optically thick BLR based on the bowl-shaped torus model.  

\item We investigate the physical properties of the outflow and find that the outflow may help to form and maintain the hot dust component. This finding is consistent with the ``radiation-driven fountain model'', in which the wind and outflow from the accretion disk drive the dusty torus.


\item Our results are somewhat inconsistent with previous studies based on the luminous AGNs, revealing that the sample spanning a wide range of AGN properties is vital to minimize the selection bias.
\end{itemize}

\begin{acknowledgments}
We are grateful to the anonymous referee for constructive comments and suggestions that greatly helped us improve our manuscript. LCH was supported by the National Science Foundation of China (11721303, 11991052, 12011540375, 12233001) and the China Manned Space Project (CMS-CSST-2021-A04, CMS-CSST-2021-A06). This work was supported by the National Research Foundation of Korea (NRF) grant funded by the Korean government (MSIT) (No. 2022R1A4A3031306 and 2023R1A2C1006261).
\end{acknowledgments}

\medskip

\facilities{IRSA}

\smallskip

\software{Astropy \citep{astropy_2013,astropy_2018,astropy_2022}, Scipy \citep{scipy_2020}, LePHARE \citep{arnouts_2011}}

\bibliography{ms}

\restartappendixnumbering
\appendix
\section{Examples of Poor Fits}
Here, we present representative examples of poor fits with a reduced $\chi^2$-value greater than 50. Note that the sample with poor fits can be categorized into three categories. First, disagreement at the shorter wavelengths (e.g., $u$ band) is occasionally detected, revealing that the featureless continuum from the accretion disk is not perfectly modeled (Fig. A1). Second, some of the AGNs with poor fits exhibit substantial excess in the MIR, possibly due to the enhancement of the hot dust component or the contribution from cold dust, which is not adequately modeled with the current templates (Fig. A2). Finally, the poor fits with the largest reduced $\chi^2$ ($\ge 100$) are mostly driven by sources that are undetected in 2MASS. 2MASS provides $2\sigma$ upper limits for the nondetections, which results in significant disagreement between the observations and best-fit model (Fig. A3). It suggests that the upper limits given by 2MASS may be substantially underestimated. Note that, in the LePHARE code, model SEDs are not allowed to exceed the $3\sigma$ upper limits.

\begin{figure*}[b]
\centering
\includegraphics[width=0.99\textwidth]{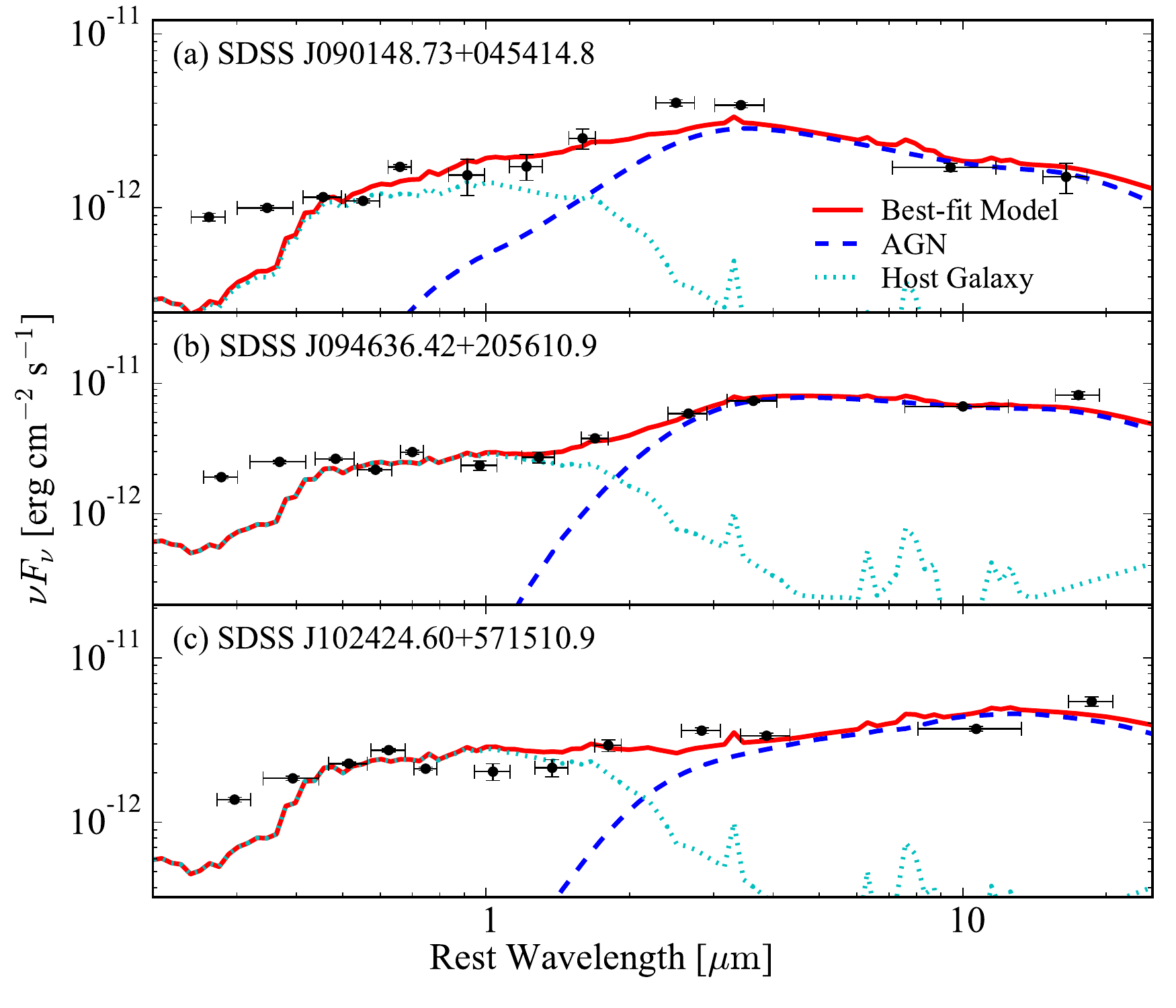}
\caption{
Representative examples of the poor fit clearly show the disagreement between the observed data and the model at the short wavelength end, such as $ug$-bands. Symbols are the same as in Figure 1.
}
\end{figure*}

\begin{figure*}[t!]
\centering
\includegraphics[width=0.99\textwidth]{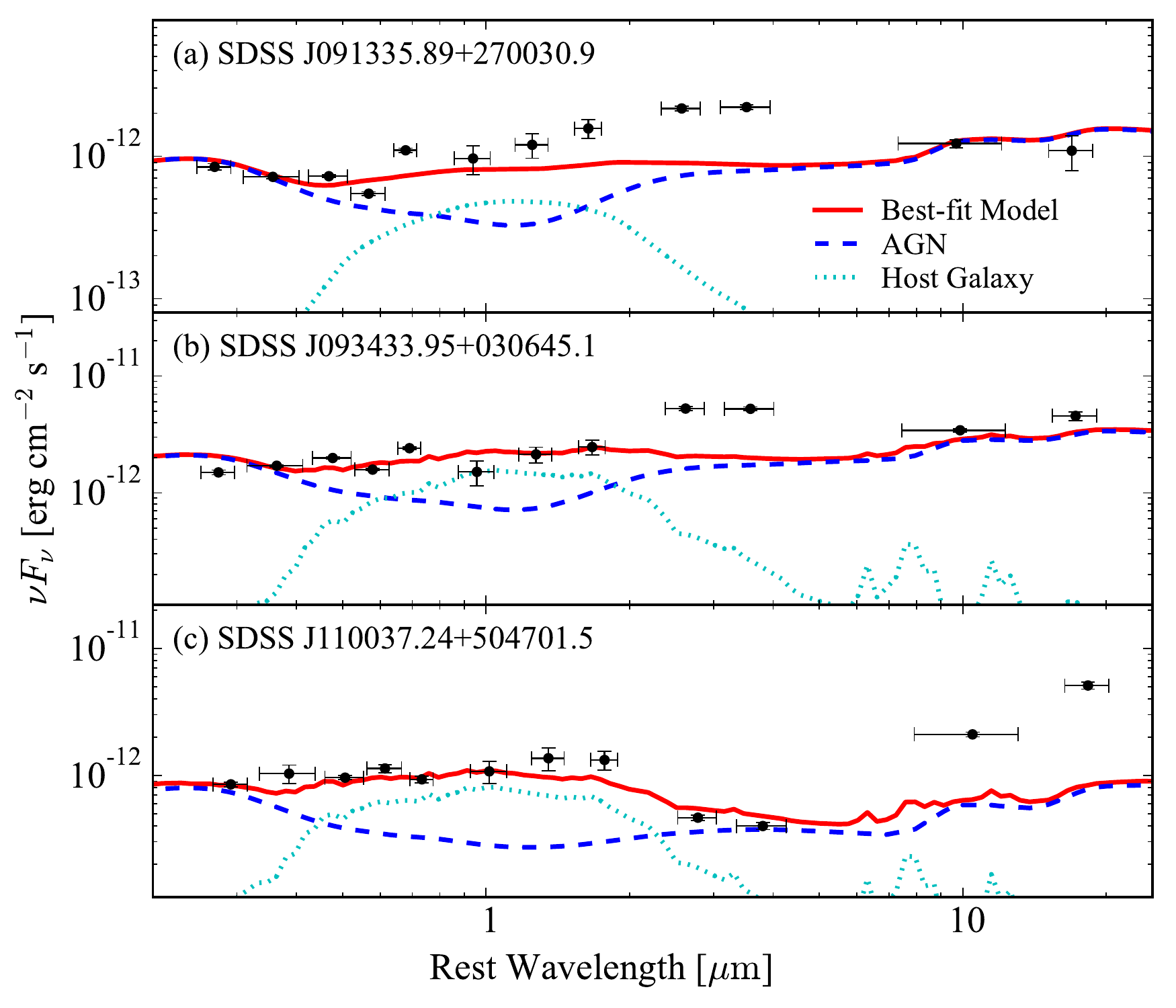}
\caption{
Representative examples of poor fits, where the residuals are substantial in MIR bands. Symbols are the same as in Figure 1.
}
\end{figure*}

\begin{figure*}[t!]
\centering
\includegraphics[width=0.99\textwidth]{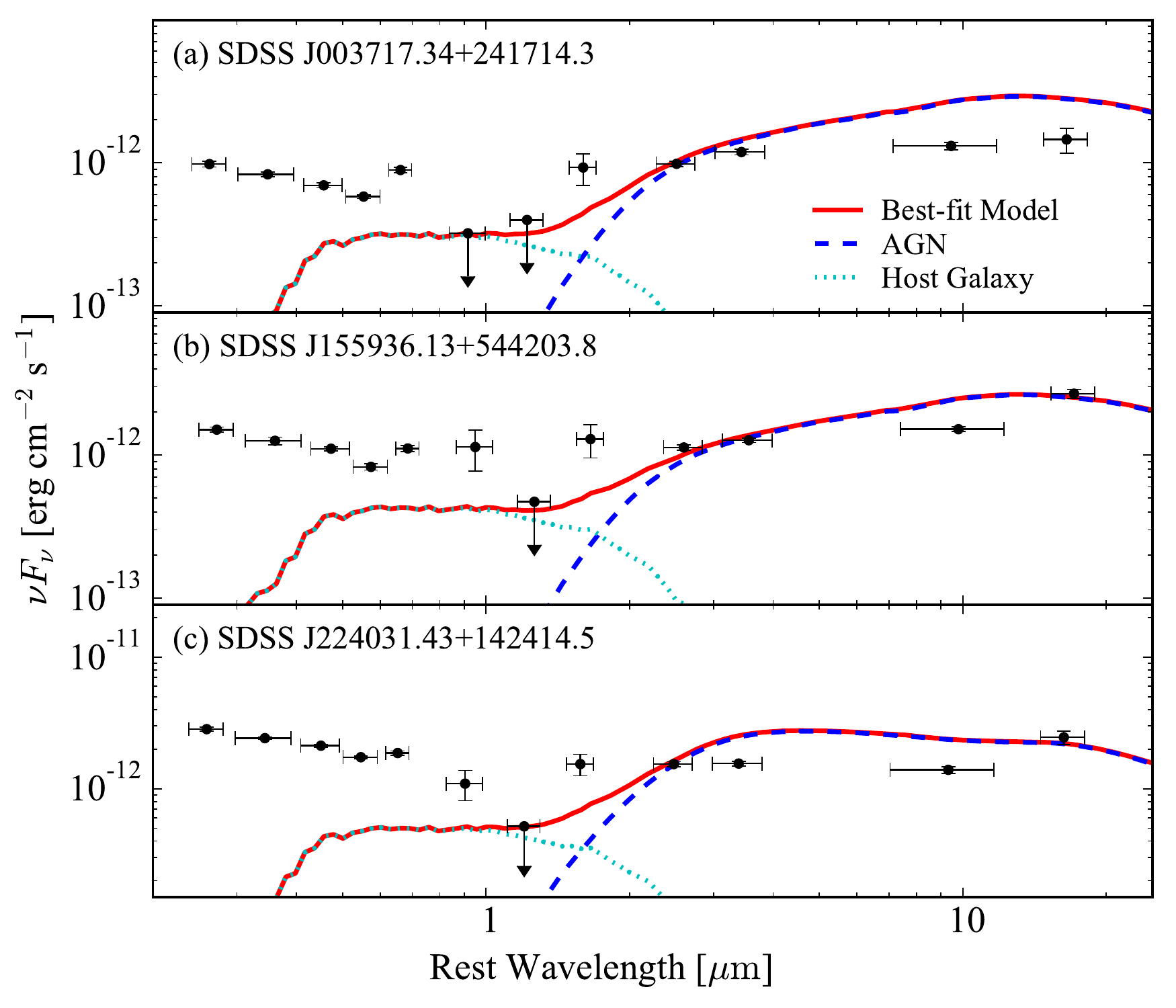}
\caption{
Representative examples of the poor fit due to underestimating the upper limit in NIR bands from 2MASS. Symbols are the same as in Figure 1.
}
\end{figure*}

\end{document}